\documentclass[runningheads]{llncs}
\usepackage[T1]{fontenc}
\usepackage{graphicx}
\usepackage{booktabs}
\usepackage[misc]{ifsym}
\newcommand{\corr}{(\Letter)}
\usepackage{mwe}
\usepackage{tabularx}
\usepackage{array} 
\usepackage[enable]{easy-todo}
\usepackage{listings}
\usepackage{xcolor}

\usepackage{url}
\usepackage{amsmath}
\usepackage{hyperref}

\def\BibTeX{{\rm B\kern-.05em{\sc i\kern-.025em b}\kern-.08em
    T\kern-.1667em\lower.7ex\hbox{E}\kern-.125emX}}

\begin{document}

\title{PIPER: Content-Based Table Search via profiling and LLM-Generated Pseudoqueries}

\titlerunning{PIPER}


\author{Riccardo Terrenzi\inst{1} \corr \and
Matteo Falconi\inst{2}  \and
Serkan Ayvaz\inst{1} \and
Pierluigi Plebani\inst{2}}


\authorrunning{R. Terrenzi et al.}

\institute{Centre for Industrial Software, University of Southern Denmark, Sønderborg, Denmark \email{\{rite, seay\}@mmmi.sdu.dk}
\and
Department of Electronics, Information and Bioengineering, Politecnico di Milano, Milano, Italy \email{\{matteo.falconi, pierluigi.plebani\}@polimi.it}}

\maketitle              

\begin{abstract}
The rapid growth of tabular datasets in data lakes, data spaces, and open data portals makes effective dataset search essential for reuse and analysis. Existing search systems rely mainly on metadata, which is often incomplete or low quality, especially for tables whose meaning depends on both schema and cell values. Recent advances in Large Language Models (LLMs) enable richer, content-based representations of tables. However, prior LLM-based retrieval methods have focused on Table Question Answering, where the goal is to select a single table to answer a question, rather than retrieve and rank relevant datasets. We propose PIPER, a content-driven retrieval method for tabular datasets that uses table profiles and LLM-generated queries embedded for dense retrieval. Designed for dataset search in poor-metadata settings, PIPER outperforms both classical metadata-based baselines and strong TableQA retrieval methods, demonstrating the value of LLM-based content modeling for tabular dataset search.
\end{abstract}

\section{Introduction}
Large-scale data ecosystems, including data lakes \cite{hai2023data}, data spaces \cite{Halevy_Franklin_Maier_2006a}, and open data portals, are increasingly used to support data sharing and analysis across organizational boundaries. As these environments grow in scale and heterogeneity, identifying relevant datasets becomes increasingly difficult. Dataset findability is therefore a fundamental requirement for data integration, analytics, and governance.

This challenge is closely related to the FAIR principles \cite{wilkinson2016fair}, which emphasize that data must be findable before it can be effectively accessed, interoperated with, and reused. In practice, however, dataset discovery still relies largely on metadata catalogs~\cite{Chapman_Simperl_Koesten_Konstantinidis_Ibáñez_Kacprzak_Groth_2020,Paton_Chen_Wu_2024,brickley2019google}, where datasets are indexed through titles, descriptions, and tags. While this paradigm is scalable and easy to deploy, it is fundamentally limited by metadata quality, which is often incomplete, inconsistent, or only weakly descriptive \cite{bogatu2020dataset}, especially in decentralized and cross-organizational settings \cite{nandi2025omnimesh}.

These limitations are particularly evident for tabular data, which constitutes a large share of structured datasets. For tables, semantics are not expressed only through textual descriptions, but also through the interaction between schema and values. Column names may be ambiguous, metadata may omit important context, and the meaning of a dataset often emerges only from its content. As a result, metadata-driven retrieval can fail to capture the actual informational capabilities of a table \cite{ji2025target,Zhang_Liu_Wei-Lun_Hung_Santos_Freire_2025}.

Recent advances in large language models (LLMs) \cite{vaswani2023attentionneed} and dense retrieval provide new opportunities for addressing this problem. Rather than relying exclusively on metadata, tabular datasets can be represented through content-derived descriptions that expose statistical properties, semantic cues, and likely information needs supported by the table. In this setting, LLMs can help transform table content into richer retrieval representations, for example by summarizing structure, interpreting column semantics, and generating synthetic natural-language expressions of what a dataset can answer \cite{sui2023gpt4table}.

Although related ideas have been explored in table retrieval for TableQA \cite{Cheng_Mao_Liu_Zhou_Li_Wang_Lin_Cao_Chen,Dong_Wang_2024}, dataset search poses a broader retrieval problem. The objective is not only to find one table that answers a question, but to retrieve candidate datasets that may support downstream analytical tasks, integration, or reuse. This requires representations that go beyond exact keyword matching and capture multiple aspects of table content.

In this paper, we propose PIPER: a content-based retrieval method for tabular dataset findability that reduces dependence on metadata by indexing tables through content-derived representations. Our approach combines profiling signals extracted from table content with LLM-generated pseudoqueries that represent the types of information a dataset can support. These representations are embedded for retrieval, while user queries can be reformulated and expanded to better match dataset content. Candidate tables are then reranked semantically to improve final retrieval quality. The goal is not to replace metadata when it is informative, but to improve dataset discovery in settings where metadata alone is insufficient.

Our contributions are:
\begin{itemize}
    \item A content-based formulation of tabular dataset search that highlights the limitations of metadata-only retrieval in heterogeneous data ecosystems.
    \item A retrieval pipeline for tabular dataset search that combines table profiling, LLM-generated pseudoqueries, query reformulation, and semantic reranking.
\end{itemize}

\section{Related Work}
\subsection{Dataset Search and LLM-based Table Retrieval}

Dataset search has traditionally been approached as keyword-based retrieval over metadata catalogs \cite{Chapman_Simperl_Koesten_Konstantinidis_Ibáñez_Kacprzak_Groth_2020}. Queries are matched against publisher-provided descriptions in settings such as Open Data Portals and web-indexed repositories \cite{brickley2019google}. This paradigm is simple and scalable, but its effectiveness is constrained by metadata quality, which is often incomplete, ambiguous, or poorly aligned with user intent. Similar limitations also arise in heterogeneous data lakes and data spaces, where discovery typically relies on metadata filtering \cite{Wang_Song_Chen_2016,liu2006answering}.

To reduce lexical mismatch, later work introduced neural retrieval models that encode queries and datasets into dense representations, including approaches such as Octopus \cite{octopus}, TRSS \cite{zhang2018ad}, and TSBERT \cite{chen2020table}. Related methods exploit schema matching, column mapping, or structural alignment signals \cite{wwt}. These approaches improve semantic matching beyond exact keyword overlap, but many still depend primarily on metadata, schema tokens, or shallow structure, which is limiting when the meaning of a dataset is carried by both tabular content and organization.

Recent advances in Large Language Models (LLMs) have strengthened table understanding in tasks such as TableQA and fact verification \cite{Dong_Wang_2024}. This has motivated table linearization strategies, specialized tabular language models, and joint query-table encoders \cite{dong2022table,li2024table,chen2020table,yin2020tabert,trabelsi2022strubert}. More recent work uses LLMs to contextualize schemas, rewrite queries, enrich metadata, and support retrieval pipelines for datasets and tables \cite{wang2021retrieving,wang2023solo,silva2024improving,fujita2024inferring,falconi2025improving,Zhang_Liu_Wei-Lun_Hung_Santos_Freire_2025,Hayashi_Sakaji_Dai_Goebel_2024,Singh_Kumar_Donaparthi_Karambelkar_2025,zhou2025table,Al-Qatf_Haque_Alsamhi_Buosi_Razzaq_Timilsina_Hawbani_Curry_2025}. These directions also include RAG-style settings. However, most of these systems either target single-table selection for downstream question answering or continue to rely mainly on metadata-level representations. In contrast, our setting treats each candidate dataset as a table and focuses on ranked dataset retrieval grounded directly in table content and structure.

\section{Methodology}
We address the problem of content-based dataset search for tabular data in settings where metadata are absent, incomplete, or unreliable. In our setting, each dataset corresponds to a single table. Given a collection of tables $\mathcal{D}=\{D_1,\dots,D_M\}$ and a natural language query $q$, the goal is to return a ranked list of relevant datasets. Unlike traditional dataset search systems, our method does not rely on metadata and operates only on dataset content.

The proposed approach is organized into two phases: (i) an offline phase, which indexes each dataset through synthetic user-oriented pseudoqueries generated from its content, and (ii) an online phase, which transforms a user query into optimized subqueries, retrieves candidate datasets by similarity search, and reranks them with an LLM. The overall intuition is to bridge the gap between how datasets are represented internally and how users express information needs in natural language.

Formally, the offline phase maps each dataset $D_i$ to a statistical profile $P_i$ and then to a set of pseudoqueries
\[
Q_i = \{s_{i1}, \dots, s_{iT}\},
\]
where $T$ is the number of synthetic questions generated per dataset. Each pseudoquery is embedded and stored in a vector database together with the identifier and profile of the originating dataset. At query time, the online phase maps the user query $q$ to a set of optimized subqueries
\[
U(q)=\{u_1,\dots,u_N\},
\]
retrieves the most similar pseudoqueries for each subquery, aggregates matches at dataset level, and reranks the resulting candidates.

This design is motivated by two assumptions. First, dataset publishers may not provide high-quality metadata, making metadata-centered retrieval brittle in practice. Second, users typically formulate information needs in natural language rather than through a controlled schema-aware query language. Our method therefore replaces metadata-dependent indexing with content-derived pseudoqueries and supports free-form user queries through LLM-based query optimization.

\subsection{Offline phase}

\begin{figure}[t]
\centerline{\includegraphics[width=\textwidth]{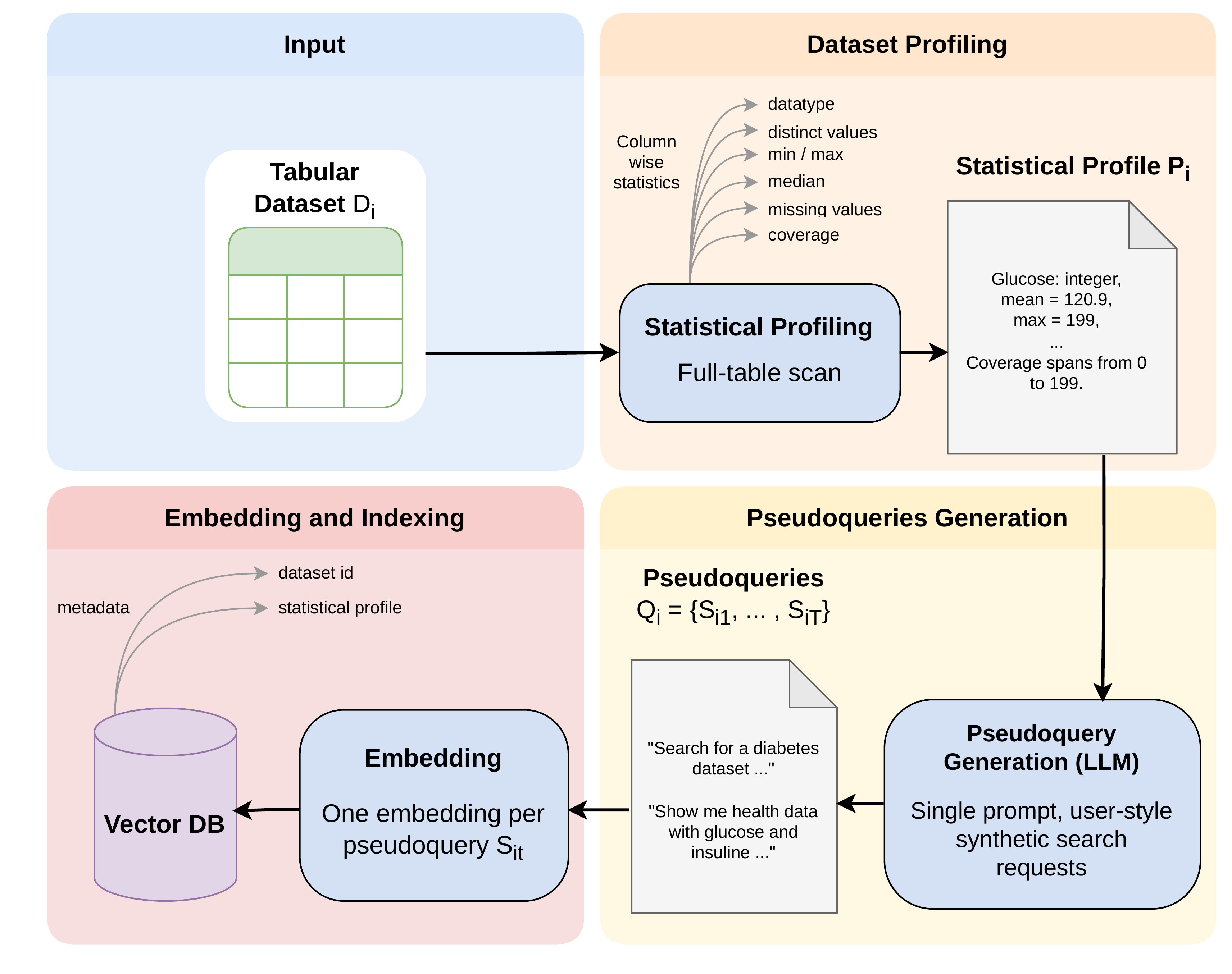}}
\caption{Architecture of the offline phase of the proposed method.}
\label{fig:offline-phase}       
\end{figure}

The offline phase transforms each raw dataset into a searchable representation derived only from its content. Its output is a set of pseudoqueries that approximate the kinds of requests a user might issue to retrieve the dataset. The structure of the offline phase is illustrated in Fig.~\ref{fig:offline-phase}.

\subsubsection{Statistical profiling}

Directly encoding large real-world tables with an LLM is often impractical due to input size constraints and the structured nature of tabular data. Rather than passing raw tables or only a small row sample, we summarize each dataset through a statistical profile computed over the full table. This avoids sampling bias and produces a compact representation of the dataset content suitable for downstream LLM processing.

Formally, let a dataset be a table
\[
D_i = \{c_{i1}, \dots, c_{im_i}\},
\]
where $m_i$ is the number of columns and each $c_{ij}$ denotes the $j$-th column of dataset $D_i$. The statistical profile of $D_i$ is defined as
\[
P_i = \mathrm{Profile}(D_i) = \{\phi(c_{i1}), \dots, \phi(c_{im_i})\},
\]
where $\phi(c_{ij})$ is a column-level summary function that maps each column to a set of descriptive statistics.

For each column, the profile includes a set of statistics that depends on the column type. In general, we compute attributes such as datatype, number of unique values, missing-value information, and value coverage. For numerical columns, the profile may additionally include statistics such as minimum, maximum, mean, and median. Hence, for a column $c_{ij}$, we can write
\[
\phi(c_{ij}) = \{t_{ij}, u_{ij}, \mu_{ij}^{\text{miss}}, \gamma_{ij}, \psi(c_{ij})\},
\]
where $t_{ij}$ denotes the detected datatype, $u_{ij}$ the number of distinct values, $\mu_{ij}^{\text{miss}}$ the missing-value information, $\gamma_{ij}$ the value coverage, and $\psi(c_{ij})$ the set of type-specific statistics. For instance, if $c_{ij}$ is numerical, then
\[
\psi(c_{ij}) = \{\min(c_{ij}), \max(c_{ij}), \mathrm{mean}(c_{ij}), \mathrm{median}(c_{ij})\}.
\]

A snippet of such a profile is shown in Listing~\ref{listing:statistical_profile}. The resulting profile does not aim to preserve cell-level detail; rather, it captures the main statistical properties of the dataset that are useful for retrieval.

\begin{lstlisting}[float, caption={Snippet of statistical profile of a single column.}, label={listing:statistical_profile}]
**Glucose**: Data is of type integer. There are 136 unique values. This column is numeric. Mean: 120.89453125, Max: 199, Min: 0. Coverage spans from 0 to 196.0.
\end{lstlisting}

\subsubsection{Pseudoquery generation}

Given the statistical profile $P_i$, we generate a fixed set of synthetic pseudoqueries that approximate the ways in which a user may search for the dataset in natural language. The goal of this step is to bridge the gap between the internal content-based representation of a table and the external linguistic formulations used at query time.

Formally, for each dataset $D_i$, the pseudoquery generation step is defined as
\[
Q_i = \mathrm{GeneratePseudoqueries}(P_i) = \{s_{i1}, \dots, s_{iT}\},
\]
where $T$ is the number of pseudoqueries generated for the dataset and each $s_{it}$ is a natural language pseudoquery intended to represent one possible user-oriented access path to $D_i$.

The generation is performed in a single LLM call. The prompt includes the statistical profile, and explicitly instructs the model to produce pseudoqueries that: (i) resemble realistic dataset search requests, (ii) cover the dataset broadly rather than focusing on a small subset of attributes, and (iii) mention relevant relationships among variables whenever these are inferable from the profile. In this way, the generated pseudoqueries are not simple reformulations of column names, but compact natural language descriptions of potentially relevant information needs associated with the dataset.

This design is motivated by the target task itself. Since users typically search for datasets by expressing requests or questions in natural language, indexing a dataset through multiple pseudoqueries reduces the mismatch between user phrasing and indexed content. Moreover, representing each dataset through several pseudoqueries allows different semantic facets of the same table to be exposed separately, which is beneficial for retrieval in heterogeneous search scenarios.

An example of synthetic questions is shown in Listing~\ref{listing:synthetic_questions}.

\begin{lstlisting}[float, caption={Example of synthetic question.}, label={listing:synthetic_questions}]
Search for a diabetes dataset with patient attributes like age and BMI.
\end{lstlisting}

\subsubsection{Embedding and indexing}

Each pseudoquery $s_{ij}$ is independently embedded using an embedding model and stored in a vector database. For each embedding, we retain as metadata the identifier of the corresponding dataset and its statistical profile. Similarity search is then performed with $L2$ similarity over the pseudoquery embeddings.

As a result, each dataset is represented not by a single vector but by a set of vectors corresponding to alternative user-oriented views of the same table. This multi-query indexing strategy increases the chances of matching heterogeneous user formulations that refer to the same dataset content.

\subsection{Online phase}

\begin{figure*}[t]
\centerline{\includegraphics[width=\textwidth]{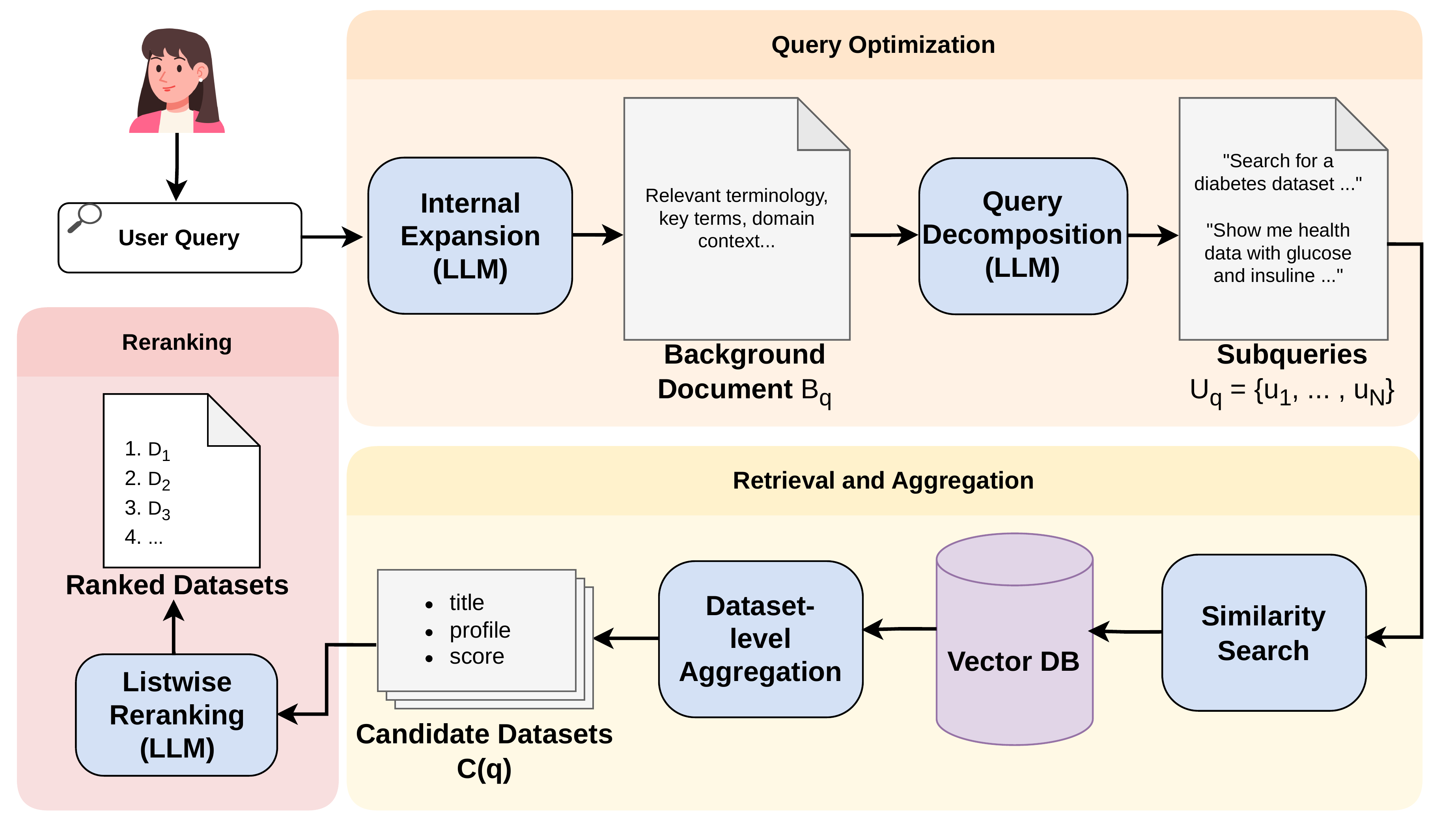}}
\caption{Architecture of the online phase of the proposed method.}
\label{fig:online-phase}
\end{figure*}

The online phase takes as input a natural language query $q$ and outputs a ranked list of datasets. It consists of three steps: query optimization, candidate retrieval, and listwise reranking. The overall structure is shown in Fig.~\ref{fig:online-phase}.

\subsubsection{Query optimization}

User queries may be incomplete, ambiguous, or phrased with terminology that does not align directly with the indexed pseudoqueries. To improve retrieval, we apply a two-step LLM-based query optimization procedure composed by (i) internal expansion and (ii) decomposition.

Given the original query $q$, the LLM first generates a short background document $B(q)$ containing contextual information, relevant terminology, and domain concepts useful for interpreting the query. The purpose of $B(q)$ is not to retrieve documents directly, but to enrich the representation of the user need before decomposition.

In a second step, the LLM receives both the original query $q$ and the background document $B(q)$ and generates a set of optimized subqueries
\[
U(q)=\{u_1,\dots,u_N\},
\]
where $N$ is dynamically determined by query complexity. These subqueries are short, explicit, and retrieval-oriented. Their role is threefold: to reduce ambiguity, to improve recall for multi-faceted information needs, and to bridge vocabulary mismatch between user phrasing and indexed pseudoqueries. An example is shown in Listing~\ref{listing:subqueries}.

\begin{lstlisting}[float, caption={Examples of subquery obtained after query optimization.}, label={listing:subqueries}]
Comprehensive diabetes datasets with 5+ years follow-up HbA1c levels treatment outcomes
\end{lstlisting}

Only the generated subqueries are used for retrieval, while the original query is preserved for the final reranking stage.

\subsubsection{Candidate retrieval}

For each optimized subquery $u_n$, we retrieve the top-$K$ most similar pseudoqueries from the vector database according to $L2$ similarity. Since each retrieved pseudoquery is associated with a dataset identifier, retrieval is performed at pseudoquery level and then aggregated at dataset level.

Let $\mathcal{R}(u_n)$ denote the top-$K$ retrieved pseudoqueries for subquery $u_n$. The candidate score of a dataset $D_i$ is computed as the number of times pseudoqueries belonging to $D_i$ appear across all retrieved sets:
\[
\mathrm{score}(D_i,q) = \sum_{n=1}^{N} \sum_{s \in \mathcal{R}(u_n)} \mathbf{1}[\mathrm{dataset}(s)=D_i].
\]
In other words, a dataset is favored when multiple pseudoqueries associated with it are matched by one or more subqueries. This simple aggregation strategy is consistent with the indexing design: datasets represented by pseudoqueries that repeatedly align with different aspects of the user request are promoted in the candidate set.

After aggregation, duplicate dataset identifiers are removed and the top candidates are retained for reranking.

\subsubsection{Listwise reranking}

The retrieval stage provides a candidate pool with coarse relevance estimates, but it does not fully exploit the richer dataset-level evidence available in the metadata associated with each pseudoquery. To refine the ranking, we apply a listwise LLM-based reranker.

Given the original query $q$ and a candidate list, where each candidate is defined as:
\[
C(q)=\{P_i, \mathrm{score}(D_i,q)\},
\]
the reranker jointly evaluates the candidate datasets and outputs a refined ranked list. For each candidate, the LLM receives the dataset identifier, its statistical profile, and the retrieval score. The original natural language query is used in this stage instead of the decomposed subqueries, so that the final ranking is aligned with the user’s initial information need rather than only with its retrieval-oriented reformulations.

The reranker is implemented in a listwise unsupervised setting with fixed candidate size and its main role is to exploit the statistical profile to distinguish datasets that may receive similar retrieval counts but differ in actual relevance to the original query.

\section{Evaluation}
\subsection{Evaluation Setup}

We evaluate the proposed method in two stages. First, we use TARGET \cite{ji2025target} as a controlled benchmark for table retrieval. Second, we evaluate on a tabular subset of NTCIR-15 Data Search \cite{Kato_Ohshima_Liu_Chen_2021} in order to assess the method in a dataset-search setting with more heterogeneous tables and queries.

For the first stage, we consider FetaQA \cite{nan2022fetaqa} and OTT-QA \cite{chen2020open}, two benchmarks included in TARGET and adapted there for table retrieval. Although our task is dataset search rather than table question answering, both settings require retrieving the relevant table from a query. FetaQA and OTT-QA involve relatively small and fairly homogeneous tables; Table~\ref{tab:dataset_stats} summarizes this contrast with the NTCIR-15 subset used in the second stage.

\begin{table}[t]
\caption{Datasets statistics for FetaQA, OTT-QA, and the NTCIR-15 tabular subset.}
\label{tab:dataset_stats}
\renewcommand{\arraystretch}{1.2}
\begin{tabularx}{\columnwidth}{l
    >{\centering\arraybackslash}X
    >{\centering\arraybackslash}X
    >{\centering\arraybackslash}X}
\toprule
\textbf{Statistic} & \textbf{FetaQA} & \textbf{OTT-QA} & \textbf{NTCIR-15 subset} \\
\midrule
\textbf{Rows (avg)} & 13.8 & 15.7 & 73.7K \\
\textbf{Cols (avg)} & 5.9 & 4.4 & 25.5 \\
\textbf{Tables} & 2K & 789 & 111 \\
\textbf{Queries} & 2K & 2.2K & 10 \\
\bottomrule
\end{tabularx}
\end{table}

For the second stage, we use NTCIR-15 Data Search, which includes heterogeneous datasets, search queries, and human relevance judgments. Since the current prototype operates directly on tabular dataset content, we restrict the evaluation to datasets that can be processed as tables. We therefore construct a subset consisting of 111 tabular datasets and 10 queries for which the majority of relevant results are tabular, so that metrics can be computed consistently within the intended retrieval setting.

On NTCIR-15, we consider two query settings: the original keyword queries provided by the benchmark and natural-language reformulations generated by an LLM. For each keyword query, the model was prompted to generate a more complex natural-language query expressing the same information need.

\subsection{Implementation Details}

Experiments on both TARGET and NTCIR-15 were conducted with the same overall setup. We used \texttt{gpt5-mini}\footnote{\url{https://developers.openai.com/api/docs/models/gpt-5-mini}} for semantic representation, pseudoquery generation, query optimization, reranking, and the generation of natural-language reformulations of the NTCIR-15 keyword queries. For both benchmarks, all generated texts were embedded with \texttt{text-embedding-3-small} and indexed in a persistent Chroma\footnote{\url{https://www.trychroma.com/}} vector store with an HNSW nearest-neighbor index using L2 distance. The same prompt-building functions were reused across datasets rather than designing benchmark-specific prompt templates. Preprocessing and type detection followed the same shared logic across benchmarks, using \texttt{datamart profiler}\footnote{\url{https://pypi.org/project/datamart-profiler/}}.

\subsection{Results on TARGET}

On TARGET, we evaluate the proposed method on FetaQA and OTT-QA using Recall, which is the metric reported in prior work for these benchmarks. Table~\ref{table:results_target} compares PIPER with the retrieval methods reported in TARGET.

\begin{table}[t]
\caption{Comparison of Recall@10 of PIPER and methods proposed in TARGET and QGpT on FetaQA and OTT-QA.}
\label{table:results_target}
\renewcommand{\arraystretch}{1.2}
\begin{tabularx}{\columnwidth}{
l 
>{\centering\arraybackslash}X 
>{\centering\arraybackslash}X}
\toprule
\textbf{Method} & \multicolumn{2}{c}{\textbf{Benchmark}} \\
\cmidrule(lr){2-3}
 & \textbf{\textit{OTT-QA}} & \textbf{\textit{FETAQA}} \\
\midrule
\textbf{Sparse Lexical Repr. (BM25) \cite{ji2025target}} & \textbf{0.967} & 0.082 \\
\textit{w/o table title} & 0.592 & 0.084 \\

\textbf{Sparse Lexical Repr. (TF-IDF) \cite{ji2025target}} & \underline{0.963} & 0.083 \\
\textit{w/o table title} & 0.583 & 0.039 \\

\textbf{Dense Metadata Embedding \cite{ji2025target}} & 0.820 & 0.436 \\
\textbf{Dense Table Embedding \cite{ji2025target}} & \underline{0.963} & 0.741 \\
\textit{column names only} & 0.658 & 0.208 \\

\textbf{Dense Row-level Embedding \cite{ji2025target}} & 0.951 & 0.711 \\

\textbf{pT \cite{liang-etal-2025-improving-table}} & 0.867 & 0.573 \\
\textbf{pT + QGpT \cite{liang-etal-2025-improving-table}} & 0.915 & 0.586 \\

\textbf{PIPER} & 0.729 & \textbf{0.784} \\
\textit{w/o query optimization} & 0.780 & \underline{0.783} \\
\bottomrule
\end{tabularx}
\end{table}

We also compare PIPER with QGpT \cite{liang-etal-2025-improving-table}, which uses synthetic questions for retrieval. For this comparison, we use the same benchmark setting and the results reported in the original QGpT work. Table~\ref{table:results_merged} reports Recall@10, Recall@5, and Recall@1 for both approaches.

\begin{table*}[t]
\caption{Recall on \textbf{FetaQA} and \textbf{OTT-QA}. Best results are in bold; second best are underlined.}
\label{table:results_merged}
\centering
\small
\setlength{\tabcolsep}{4pt}
\renewcommand{\arraystretch}{1.05}
\begin{tabular*}{\textwidth}{@{\extracolsep{\fill}}lcccccc@{}}
\toprule
& \multicolumn{3}{c}{\textbf{FetaQA}} & \multicolumn{3}{c}{\textbf{OTT-QA}} \\
\cmidrule(lr){2-4}\cmidrule(lr){5-7}
\textbf{Method} & \textit{R@10} & \textit{R@5} & \textit{R@1} & \textit{R@10} & \textit{R@5} & \textit{R@1} \\
\midrule
\textbf{pT}\cite{liang-etal-2025-improving-table}         & .573 & .508 & .353 & \underline{.867} & \underline{.781} & \underline{.515} \\
\textbf{pT + QGpT}\cite{liang-etal-2025-improving-table}  & .586 & .522 & .372 & \textbf{.915} & \textbf{.844} & \textbf{.608} \\
\textbf{PIPER}      & \textbf{.784} & \textbf{.660} & \textbf{.401} & .729 & .619 & .336 \\
\textit{w/o q. opt.} & \underline{.783} & \underline{.654} & \underline{.395} & .780 & .648 & .356 \\
\bottomrule
\end{tabular*}
\end{table*}

\subsection{Results on NTCIR-15 Data Search}

We next evaluate PIPER on the tabular subset of NTCIR-15. All compared methods were run on the same subset in order to ensure a fair comparison. We report MAP, Precision@10, Recall@10, and nDCG@10 for both the original keyword queries and the \texttt{gpt-5-mini}-generated natural-language reformulations. The results are shown in Table~\ref{table:results}.

\begin{table*}[t]
\caption{Results on NTCIR-15 for PIPER and baseline retrieval models. Best results are in bold; second best are underlined.}
\label{table:results}
\centering
\small
\setlength{\tabcolsep}{4pt}
\renewcommand{\arraystretch}{1.05}
\begin{tabular*}{\textwidth}{@{\extracolsep{\fill}}lcccccccc@{}}
\toprule
& \multicolumn{4}{c}{\textbf{Complex NL}} & \multicolumn{4}{c}{\textbf{Keyword}} \\
\cmidrule(lr){2-5}\cmidrule(lr){6-9}
\textbf{Model} & \textit{MAP} & \textit{P@10} & \textit{R@10} & \textit{nDCG@10} 
               & \textit{MAP} & \textit{P@10} & \textit{R@10} & \textit{nDCG@10} \\
\midrule
\textbf{TAPAS-base}         & .035 & .110 & .104 & .128 & .089 & .160 & .161 & .177 \\
\textbf{BM25}               & .192 & .250 & .347 & .336 & .204 & .260 & .301 & .330 \\
\textbf{SPLADE}             & .234 & .280 & .373 & .386 & .253 & .270 & .339 & .379 \\
\textbf{ColBERTv2}          & .242 & .280 & .389 & .389 & .263 & .270 & .369 & .398 \\
\textbf{Dense-BGE}         & \underline{.364} & \underline{.360} & \underline{.468} & \underline{.510}
                   & \underline{.342} & \underline{.360} & \underline{.480} & \underline{.487} \\
\textbf{PIPER}              & \textbf{.560} & \textbf{.480} & \textbf{.647} & \textbf{.676}
                   & \textbf{.483} & \textbf{.430} & \textbf{.563} & \textbf{.578} \\
\textit{w/o q. opt.} & .161 & .150 & .188 & .233 & .200 & .150 & .220 & .327 \\
\bottomrule
\end{tabular*}
\end{table*}

\begin{figure*}[t]
\centerline{\includegraphics[width=0.8\textwidth]{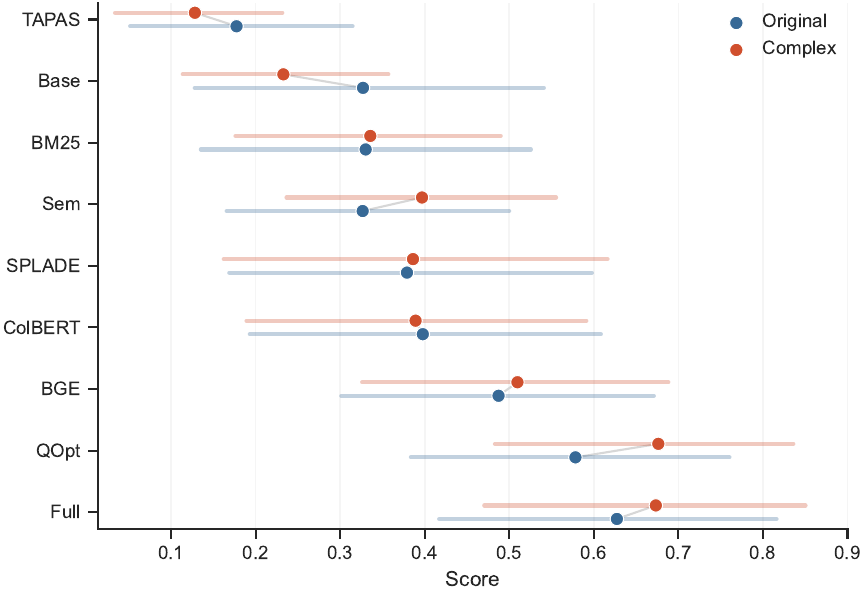}}
\caption{95\% bootstrap confidence intervals for nDCG@10 on the NTCIR-15 tabular subset. Full stands for full PIPER system. QOpt stands for no query optimization ablation.}
\label{fig:bootstrap}
\end{figure*}

For NTCIR-15, we additionally computed 95\% bootstrap confidence intervals for the evaluation metrics. Figure~\ref{fig:bootstrap} reports the corresponding intervals for nDCG@10.

\section{Discussion}
\paragraph{Sensitivity to metadata quality.}
The results suggest that the proposed method is most useful in retrieval settings where metadata is weak, incomplete, or poorly aligned with the query. This pattern is visible in FetaQA and in the NTCIR-15 subset, where retrieval must rely more directly on dataset content and where our approach is consistently competitive or superior. By contrast, on OTT-QA, where titles and metadata provide strong lexical cues, metadata-aware methods remain stronger. In our view, this is not a contradiction but a clear indication of the role of the method: it should be understood primarily as a content-based alternative for metadata-poor environments, rather than as a universal replacement for metadata-driven retrieval.

\paragraph{Profiles instead of partial table views.}
The comparison with QGpT also highlights an important design choice. Both approaches use synthetic questions, but QGpT benefits from metadata and partial table views, whereas our method relies on semantic profiles derived from full dataset content. The results indicate that profile-based representations can be more robust when metadata is unreliable, since they capture broader signals about what a dataset is about rather than depending on a few exposed fields or textual descriptors. At the same time, the stronger OTT-QA performance of QGpT suggests that partial-table and metadata-based signals remain highly effective when those signals are informative. This points to complementarity rather than clear dominance of one strategy over the other.

\paragraph{When query optimization helps.}
The ablation results suggest that query optimization is not uniformly beneficial. On TARGET, where queries are already tightly aligned with the target tables, its effect is small and can even be slightly negative, likely because additional reformulation introduces mild query drift. In contrast, the stronger results on natural-language queries in NTCIR-15 suggest that query optimization becomes more useful when information needs are expressed more freely and lexical overlap with the relevant dataset is less direct. We therefore see query optimization as an adaptive component: valuable in realistic dataset search, but less necessary in tightly controlled benchmark settings.

\paragraph{Implications.}
Overall, the discussion points to a simple conclusion: dataset retrieval is not governed by a single best representation, but by the interaction between query formulation, metadata quality, and how much of the dataset content is actually exposed to the retriever. The main strength of the proposed method is that it shifts retrieval toward content-derived semantic evidence, which is particularly important in real collections where metadata cannot be assumed to be sufficient. At the same time, the results also make clear that informative metadata should not be ignored. For this reason, the most promising direction for future work is not a stronger content-only model, but hybrid retrieval strategies that can adaptively combine metadata, semantic profiles, and synthetic questions depending on the characteristics of the collection.

\section{Conclusion}
This paper introduced a profiling-based indexing method for table and dataset search in which table summaries are used to guide the generation of LLM-based pseudoqueries for content-driven retrieval. The method is intended for settings where user needs are expressed in natural language and where metadata alone is insufficient to support effective discovery. The experimental results on TARGET show competitive performance against established baselines, especially in metadata-poor scenarios, while the evaluation on a subset of NTCIR-15 indicates that the approach remains effective in a more realistic dataset-search setting. These findings highlight the value of integrating table understanding and retrieval to improve tabular data findability. They also suggest that the most effective search systems will likely combine metadata-aware and content-driven retrieval, with future work needed on adaptive strategies, efficiency, and broader evaluation across heterogeneous collections.


\begin{thebibliography}{99}
\providecommand{\url}[1]{\texttt{#1}}
\providecommand{\urlprefix}{URL }
\providecommand{\doi}[1]{https://doi.org/#1}

\bibitem{Halevy_Franklin_Maier_2006a}
Halevy, A., Franklin, M., Maier, D.: Principles of dataspace systems. In: Proceedings of the
  twenty-fifth ACM SIGMOD-SIGACT-SIGART symposium on Principles of database
  systems. p. 1–9. ACM, Chicago IL USA (Jun 2006).
  \doi{10.1145/1142351.1142352},
  \url{https://dl.acm.org/doi/10.1145/1142351.1142352}

\bibitem{Kato_Ohshima_Liu_Chen_2021}
Kato, M.P., Ohshima, H., Liu, Y.-H., Chen, H.-L.: A test collection for ad-hoc dataset retrieval. In: Proceedings of
  the 44th International ACM SIGIR Conference on Research and Development in
  Information Retrieval. p. 2450–2456. ACM, Virtual Event Canada (Jul 2021).
  \doi{10.1145/3404835.3463261},
  \url{https://dl.acm.org/doi/10.1145/3404835.3463261}

\bibitem{Zhang_Liu_Wei-Lun_Hung_Santos_Freire_2025}
Zhang, H., Liu, Y., Hung, W.-L., Santos, A., Freire, J.: Autoddg: Automated dataset description generation using large
  language models (arXiv:2502.01050) (Feb 2025).
  \doi{10.48550/arXiv.2502.01050}, \url{http://arxiv.org/abs/2502.01050},
  arXiv:2502.01050 [cs]

\bibitem{Al-Qatf_Haque_Alsamhi_Buosi_Razzaq_Timilsina_Hawbani_Curry_2025}
Al-Qatf, M., Haque, R., Alsamhi, S.H., Buosi, S., Razzaq, M.A., Timilsina, M.,
  Hawbani, A., Curry, E.: {RAG4DS: Retrieval-Augmented Generation for Data
  Spaces—A Unified Lifecycle, Challenges, and Opportunities}. IEEE Access
  \textbf{13},  39510–39522 (2025). \doi{10.1109/ACCESS.2025.3545387}

\bibitem{bogatu2020dataset}
Bogatu, A., Fernandes, A.A., Paton, N.W., Konstantinou, N.: Dataset discovery
  in data lakes. In: 2020 IEEE 36th international Conference on Data
  Engineering (ICDE). pp. 709--720. IEEE (2020)

\bibitem{brickley2019google}
Brickley, D., Burgess, M., Noy, N.: Google dataset search: Building a search
  engine for datasets in an open web ecosystem. In: The World Wide Web
  Conference. p. 1365–1375. WWW '19, Association for Computing Machinery, New
  York, NY, USA (2019). \doi{10.1145/3308558.3313685},
  \url{https://doi.org/10.1145/3308558.3313685}

\bibitem{octopus}
Cafarella, M.J., Halevy, A., Khoussainova, N.: Data integration for the
  relational web  \textbf{2}(1) (2009). \doi{10.14778/1687627.1687750}

\bibitem{Chapman_Simperl_Koesten_Konstantinidis_Ibáñez_Kacprzak_Groth_2020}
Chapman, A., Simperl, E., Koesten, L., Konstantinidis, G., Ibáñez, L.D.,
  Kacprzak, E., Groth, P.: Dataset search: a survey. The VLDB Journal
  \textbf{29}(1),  251–272 (Jan 2020). \doi{10.1007/s00778-019-00564-x}

\bibitem{chen2020open}
Chen, W., Chang, M.W., Schlinger, E., Wang, W., Cohen, W.W.: Open question
  answering over tables and text. arXiv preprint arXiv:2010.10439  (2020)

\bibitem{chen2020table}
Chen, Z., Trabelsi, M., Heflin, J., Xu, Y., Davison, B.D.: Table search using a
  deep contextualized language model. In: Proceedings of the 43rd International
  ACM SIGIR Conference on Research and Development in Information Retrieval.
  pp. 589--598 (2020)

\bibitem{Cheng_Mao_Liu_Zhou_Li_Wang_Lin_Cao_Chen}
Cheng, M., Mao, Q., Liu, Q., Zhou, Y., Li, Y., Wang, J., Lin, J., Cao, J.,
  Chen, E.: A survey on table mining with large language models: Challenges,
  advancements and prospects. TechRxiv  (April 2025).
  \doi{10.36227/techrxiv.174352282.22844759/v1}

\bibitem{dong2022table}
Dong, H., Cheng, Z., He, X., Zhou, M., Zhou, A., Zhou, F., Liu, A., Han, S.,
  Zhang, D.: Table pre-training: A survey on model architectures, pre-training
  objectives, and downstream tasks. arXiv preprint arXiv:2201.09745  (2022)

\bibitem{Dong_Wang_2024}
Dong, H., Wang, Z.: Large language models for tabular data: Progresses and
  future directions. In: Proceedings of the 47th International ACM SIGIR
  Conference on Research and Development in Information Retrieval. p.
  2997–3000. SIGIR ’24, Association for Computing Machinery, New York, NY,
  USA (2024). \doi{10.1145/3626772.3661384},
  \url{https://dl.acm.org/doi/10.1145/3626772.3661384}

\bibitem{falconi2025improving}
Falconi, M., Plebani, P.: Improving content-based data product retrieval in
  federated environments with llm and sampling. In: International Conference on
  Advanced Information Systems Engineering. pp. 289--297. Springer (2025)

\bibitem{fujita2024inferring}
Fujita, Y., Hayashi, T., Kuwahara, M.: Inferring relationships between tabular
  data and topics using llm for a dataset search task. In: 2024 IEEE
  International Conference on Big Data (BigData). pp. 6564--6573. IEEE (2024)

\bibitem{hai2023data}
Hai, R., Koutras, C., Quix, C., Jarke, M.: Data lakes: A survey of functions
  and systems. IEEE Transactions on Knowledge and Data Engineering
  \textbf{35}(12),  12571--12590 (2023)

\bibitem{Hayashi_Sakaji_Dai_Goebel_2024}
Hayashi, T., Sakaji, H., Dai, J., Goebel, R.: Metadata-based data exploration
  with retrieval-augmented generation for large language models. In: 2024 IEEE
  International Conference on Big Data (BigData). p. 6574–6583 (Dec 2024).
  \doi{10.1109/BigData62323.2024.10826055},
  \url{https://ieeexplore.ieee.org/abstract/document/10826055}

\bibitem{ji2025target}
Ji, X., Glenn, P., Parameswaran, A.G., Hulsebos, M.: Target: Benchmarking table
  retrieval for generative tasks. arXiv preprint arXiv:2505.11545  (2025)

\bibitem{li2024table}
Li, P., He, Y., Yashar, D., Cui, W., Ge, S., Zhang, H., Rifinski~Fainman, D.,
  Zhang, D., Chaudhuri, S.: Table-gpt: Table fine-tuned gpt for diverse table
  tasks. Proceedings of the ACM on Management of Data  \textbf{2}(3),  1--28
  (2024)

\bibitem{liang-etal-2025-improving-table}
Liang, H.P., Chang, C.W., Fan, Y.C.: Improving table retrieval with question
  generation from partial tables. In: Chang, S., Hulsebos, M., Liu, Q., Chen,
  W., Sun, H. (eds.) Proceedings of the 4th Table Representation Learning
  Workshop. pp. 217--228. Association for Computational Linguistics, Vienna,
  Austria (Jul 2025). \doi{10.18653/v1/2025.trl-1.19},
  \url{https://aclanthology.org/2025.trl-1.19/}

\bibitem{liu2006answering}
Liu, J., Dong, X., Halevy, A.Y.: Answering structured queries on unstructured
  data. In: WebDB. vol.~6, pp. 25--30 (2006)

\bibitem{nan2022fetaqa}
Nan, L., Hsieh, C., Mao, Z., Lin, X.V., Verma, N., Zhang, R., Kry{\'s}ci{\'n}ski, W., Schoelkopf, H., Kong, R., Tang, X., Mutuma, M., Rosand, B., Trindade, I., Bandaru, R., Cunningham, J., Xiong, C., Radev, D.: Fetaqa:
  Free-form table question answering. Transactions of the Association for
  Computational Linguistics  \textbf{10},  35--49 (2022)

\bibitem{nandi2025omnimesh}
Nandi, A., Chao, W.L., Qin, R., Boettiger, C., Lapp, H., Berger-Wolf, T.:
  Omnimesh: Addressing findability challenges in distributed nature data
  repositories. In: Proceedings of the 37th International Conference on
  Scalable Scientific Data Management. pp.~1--6 (2025)

\bibitem{Paton_Chen_Wu_2024}
Paton, N.W., Chen, J., Wu, Z.: Dataset discovery and exploration: A survey. ACM
  Computing Surveys  \textbf{56}(4),  1–37 (Apr 2024). \doi{10.1145/3626521}

\bibitem{wwt}
Pimplikar, R., Sarawagi, S.: Answering table queries on the web using column
  keywords. arXiv preprint arXiv:1207.0132  (2012)

\bibitem{silva2024improving}
Silva, L., Barbosa, L.: {Improving dense retrieval models with LLM augmented
  data for dataset search}. Knowledge-based systems  \textbf{294},  111740
  (2024)

\bibitem{Singh_Kumar_Donaparthi_Karambelkar_2025}
Singh, M., Kumar, A., Donaparthi, S., Karambelkar, G.: {Leveraging Retrieval
  Augmented Generative LLMs For Automated Metadata Description Generation to
  Enhance Data Catalogs} (arXiv:2503.09003) (Mar 2025).
  \doi{10.48550/arXiv.2503.09003}, \url{http://arxiv.org/abs/2503.09003}

\bibitem{sui2023gpt4table}
Sui, Y., Zhou, M., Zhou, M., Han, S., Zhang, D.: Gpt4table: Can large language
  models understand structured table data? a benchmark and empirical study.
  arXiv preprint ArXiv:2305.13062  (2023)

\bibitem{trabelsi2022strubert}
Trabelsi, M., Chen, Z., Zhang, S., Davison, B.D., Heflin, J.: Strubert:
  Structure-aware bert for table search and matching. In: Proceedings of the
  ACM Web Conference 2022. pp. 442--451 (2022)

\bibitem{vaswani2023attentionneed}
Vaswani, A., Shazeer, N., Parmar, N., Uszkoreit, J., Jones, L., Gomez, A.N.,
  Kaiser, L., Polosukhin, I.: {Attention Is All You Need} (arXiv:1706.03762)
  (Aug 2023). \doi{10.48550/arXiv.1706.03762},
  \url{http://arxiv.org/abs/1706.03762}

\bibitem{wang2021retrieving}
Wang, F., Sun, K., Chen, M., Pujara, J., Szekely, P.: Retrieving complex tables
  with multi-granular graph representation learning. In: Proceedings of the
  44th International ACM SIGIR Conference on Research and Development in
  Information Retrieval. pp. 1472--1482 (2021)

\bibitem{wang2023solo}
Wang, Q., Castro~Fernandez, R.: Solo: Data discovery using natural language
  questions via a self-supervised approach. Proceedings of the ACM on
  Management of Data  \textbf{1}(4),  1--27 (2023)

\bibitem{Wang_Song_Chen_2016}
Wang, Y., Song, S., Chen, L.: A survey on accessing dataspaces. ACM SIGMOD
  Record  \textbf{45}(2),  33–44 (Sep 2016). \doi{10.1145/3003665.3003672}

\bibitem{wilkinson2016fair}
Wilkinson, M.D., Dumontier, M., Aalbersberg, I.J., Appleton, G., Axton, M., Baak, A., Blomberg, N., Boiten, J.-W., da~Silva~Santos, L.B., Bourne, P.E., Bouwman, J., Brookes, A.J., Clark, T., Crosas, M., Dillo, I., Dumon, O., Edmunds, S., Evelo, C.T., Finkers, R., Gonz{\'a}lez-Beltr{\'a}n, A., Gray, A.J.G., Groth, P., Goble, C., Grethe, J.S., Heringa, J., {'t} Hoen, P.-B., Hooft, R., Kuhn, T., Kok, R., Kok, J., Lusher, S.J., Martone, M.E., Mons, A., Packer, A.L., Persson, B., Rocca-Serra, P., Roos, M., van~Schaik, R., Sansone, S.-A., Schultes, E., Sengstag, T., Slater, T., Strawn, G., Swertz, M.A., Thompson, M., van~der~Lei, J., van~Mulligen, E.M., Velterop, J., Waagmeester, A., Wittenburg, P., Wolstencroft, K., Zhao, J., Mons, B.: {The FAIR Guiding Principles for scientific data management and
  stewardship}. Scientific data  \textbf{3}(1), ~1--9 (2016)

\bibitem{yin2020tabert}
Yin, P., Neubig, G., Yih, W.t., Riedel, S.: {TaBERT: Pretraining for joint
  understanding of textual and tabular data}. arXiv preprint arXiv:2005.08314
  (2020)

\bibitem{zhang2018ad}
Zhang, S., Balog, K.: Ad hoc table retrieval using semantic similarity. In:
  Proceedings of the 2018 world wide web conference. pp. 1553--1562 (2018)

\bibitem{zhou2025table}
Zhou, W., Ma, B., Friedrich, A., Mesgar, M.: Table question answering in the
  era of large language models: A comprehensive survey of tasks, methods, and
  evaluation. arXiv preprint arXiv:2510.09671  (2025)

\end{thebibliography}
\end{document}